\documentstyle[twocolumn,aps,overcite,nature,bm,psfig]{revtex}  
\newif\iffigs
\figstrue   
\iffigs
\fi
\def\drawing #1 #2 #3 {
\begin{center}
\setlength{\unitlength}{1mm}
\begin{picture}(#1,#2)(0,0)
\put(0,0){\framebox(#1,#2){#3}}
\end{picture}
\end{center} }

\def\rf#1{(\ref{#1})}

\def\q{{\bf q}}
\def\x{{\bf x}}

\begin{document}
\noindent{\Large\bf  A reconstruction of the initial}

\vspace*{1.5mm}
\noindent{\Large\bf   conditions of the Universe by}

\vspace*{1.5mm}
\noindent{\Large\bf  optimal mass transportation}
\vspace*{4mm} \par\noindent {\large\bf Uriel\,\,\,\,Frisch$^*$,\,\,\,\,\,\,\,Sabino\,\,\,\,Matarrese$^{\dagger}$, Roya~Mohayaee$^{\ddag *}$ \&
Andrei~Sobolevski$^{\S *}$}
\vspace*{3mm}
\par\noindent $^*$\,\hbox{CNRS, UMR 6529, Observatoire de la C\^ote d'Azur},\\ 
\hbox{BP 4229, 06304 Nice Cedex 4, France}
\par\noindent $^\dagger$\,\hbox{Dipartimento di Fisica ``G. Galilei'' 
and INFN, Sezione} 
\par\noindent di Padova, via Marzolo 8, 35131-Padova, Italy
\par\noindent $^\ddag$\,\hbox{Dipartimento di Fisica, Universit\'a 
Degli Studi  di} 
\par\noindent  Roma ``La Sapienza'', P.le A. Moro 5, 00185-Roma, Italy
\par\noindent $^\S$\,\hbox{Department of Physics,  M.V.~Lomonossov University}
\par\noindent \hbox{119899-Moscow, Russia}

\vspace*{1mm}
\centerline{Nature$|$ VOL 417$|$ 16 MAY 2002}
\vspace{2mm} {\bf \noindent Reconstructing the density fluctuations in the
early Universe that evolved into the distribution of galaxies we see today is
a challenge of modern cosmology \cite{nacro99}.  An accurate reconstruction
would allow us to test cosmological models by simulating the evolution
starting from the reconstructed state and comparing it to the observations.
Several reconstruction techniques have been proposed
\cite{peebles89,weinberg92,ND92,CG97,NB00,GS00,VST00,BEN01}, but they all
suffer from lack of uniqueness because the velocities of galaxies are usually
not known. Here we show that reconstruction can be reduced to a
well-determined problem of optimisation, and present a specific algorithm that
provides excellent agreement when tested against data from N-body
simulations. By applying our algorithm to the new redshift surveys now under
way \cite{survey}, we will be able to recover reliably the properties of the
primeval fluctuation field of the local Universe and to determine accurately
the peculiar velocities (deviations from the Hubble expansion) and the true
positions of many more galaxies than is 
feasible by any other method.}

\begin{figure}
\iffigs 
\centerline{\psfig{file=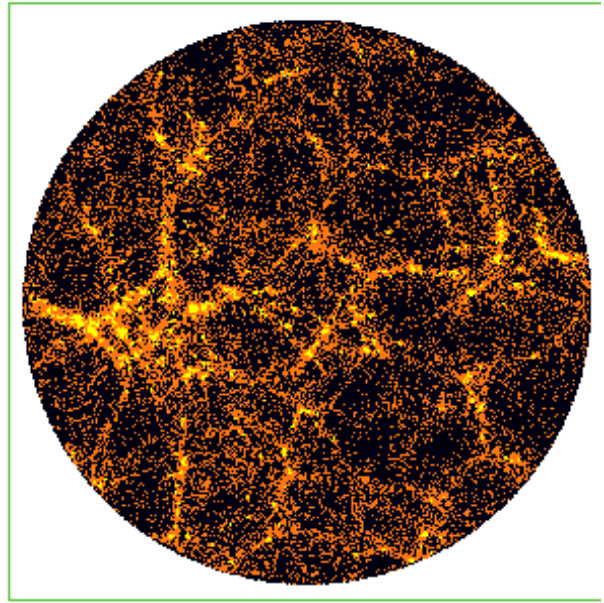,width=8cm}}

\else\drawing 65 10 {structures}
\fi
\vspace{2mm}
\caption{$N$-body simulation output (present epoch) used for testing our
reconstruction method.  In the standard model of structure formation, the
distribution of matter in the Universe is believed to have emerged from a very
smooth initial state: tiny irregularities of the gravitational potential,
which we can still observe as temperature fluctuations of the cosmic microwave
background, gave rise to density fluctuations, which grew under their
self-gravity, developing a rich and coherent pattern of structures. Most of
the mass is in the form of cold dark matter; the luminous matter (galaxies)
can be assumed to trace -- up to some form of bias -- the underlying dark
matter.  Shown here is a projection onto the $x$-$y$ plane of a 10\% slice of
the simulation box of size $200h^{-1}$\,Mpc.  The model, $\Lambda$CDM, uses
cold dark matter with cosmological constant and the following parameters:
Hubble constant $ h=0.65$, density parameters $\Omega_\Lambda=0.7$ and
$\Omega_{\rm m}=0.3$, normalization factor $\sigma_8=0.99$. Points are
highlighted in yellow when reconstruction fails by more than $6\,h^{-1}$\,Mpc,
which happens mostly in high-density regions.}
\label{f:structure}
\end{figure}

Starting from the available data on the galaxy distribution, can we trace back
in time and map to its initial locations the highly structured distribution of
mass in the Universe (Fig.~\ref{f:structure})? 
Here we show that, with a suitable 
hypothesis, the knowledge of both
the present non-uniform distribution of mass and of its primordial
quasi-uniform distribution uniquely determines the {\it inverse Lagrangian
map\/}, defined as the transformation from present (Eulerian) positions $\x$
to the respective initial (Lagrangian) positions~$\q$.

We first consider the direct Lagrangian map $\q \mapsto \x$, which can be
approximately written in terms of a potential as $\x = \nabla_\q\Phi(\q)$ at
those scales where nonlinearity stays moderate\cite{BD89}.  This is 
supported by numerical N-body simulations showing good agreement with a very
simple potential approximation, due to Zel'dovich\cite{zeldovich70}, which
assumes that the particles move on straight trajectories. Even better
agreement is obtained with a refinement, the second-order Lagrangian
perturbation method\cite{MABPR91,buchert92,MSS94,catelan95}, also known to be
potential.
In our ``reconstruction hypothesis'', we furthermore assume the convexity 
of the potential $\Phi(\q)$, a consequence of which is
the absence of multi-streaming: for almost any Eulerian position,
there is a single Lagrangian antecedent.  As is well-known, the Zel'dovich
approximation leads to caustics and to multi-streaming. This can be overcome
in various ways, for example by a modification known as the adhesion model, an
equation of viscous pressureless gas dynamics\cite{GS84,SZ89}.  The latter,
which leads to shocks rather than caustics, is known to have a convex
potential\cite{VDFN94} and to be in better agreement with N-body simulations.
Suppression or reduction of multi-streaming requires a mechanism of momentum
exchange, such as viscosity, between neighbouring streams having different
velocities.  This is a common phenomenon in ordinary fluids, such as the flow
of air or water in our natural environment.  Dark matter is however
essentially collisionless and the usual mechanism for generating viscosity
(discovered by Maxwell) 
does not operate, so that a non-collisional mechanism involving a small-scale
gravitational instability must be invoked.

Our reconstruction hypothesis implies that the initial positions can
be obtained from the present ones by another gradient map: $\q
=\nabla_{\x}\Theta(\x)$, where $\Theta$ is a convex potential related to
$\Phi$ by a Legendre--Fenchel transform (see Methods).  We denote by $\rho_0$
the initial mass density (which can be treated as uniform) and by $\rho(\x)$
the final one. Mass conservation implies $\rho_0 d^3q = \rho(\x) d^3x$. Thus,
the ratio of final to initial density is the Jacobian of the inverse
Lagrangian map. This can be written as the following Monge--Amp\`ere
equation\cite{ampere20} for the unknown potential $\Theta$
\begin{equation}
\hbox{det}\, \left(\nabla_{x_i}\nabla_{x_j} \Theta(\x)\right) =\rho(\x)/\rho_0,
\label{MA}
\end{equation}
where `det' stands for determinant. 

We emphasize that no information about the dynamics of matter other than the
reconstruction hypothesis is needed for our method, whose degree of success
depends crucially on how well this hypothesis is satisfied. Exact
reconstruction is obtained, for example, for the Zel'dovich approximation
(before particle trajectories cross) and for adhesion-model dynamics (at
arbitrary times).
 
We note that our Monge--Amp\`ere equation for self-gravitating matter may be
viewed as a nonlinear generalisation of a Poisson equation (used for
reconstruction in ref.~\citen{ND92}), to which it reduces if particles
have moved very little from their initial positions.

It has been discovered recently that the map generated by the solution to the
Monge-Amp\`ere equation \rf{MA} is the (unique) solution to an
optimisation problem \cite{brenier87} (see also refs~\citen{GM96} 
and \citen{BB00}). This is the  `mass transportation' problem of Monge and
Kantorovich \cite{monge81,kantorovich42} in which one seeks the map $\x \mapsto
\q$ that minimises the quadratic `cost' function
\begin{equation}
I = \int_\q \rho_0 |\x-\q|^2 \,d^3q = \int_\x \rho(\x) |\x-\q|^2\,d^3x.
\label{qcost}
\end{equation}
Note that $\x=\q$ is forbidden:  as the
initial and final density fields $\rho_0$ and $\rho(\x)$ are prescribed there
is a constraint on Jacobian of the map (see Methods). 

Next, we take into account that information on the mass distribution is 
provided  in the form of $N$ discrete
particles both in simulations and when handling observational data from galaxy
surveys. The cost minimisation then becomes what is known in optimisation
theory as the assignment problem: find the unique one-to-one pairing of a set
of $N$ initial points $\q_j$'s and $N$ final points $\x_i$'s that minimises $
I_{\rm discr} =\sum_{i=1}^N |\x_i-\q_{j(i)}|^2 .  $ An immediate consequence
is that, for any subset of $k$ pairs of initial and final
points ($2\le k \le N$), the contribution of these points to the cost function
should not decrease under arbitrary permutations of initial points. This
property is known to be equivalent to having a Lagrangian map that is the
gradient of a convex function \cite{rockafellar70}.

If we restrict ourselves to interchanging just pairs ($k=2$), the map is said
to be monotonic, a condition not equivalent to minimisation of the cost
function (except in one dimension).  In ref.~\citen{CG97}, a method of
reconstruction called the Path Interchange Zel'dovich Approximation (PIZA) is
introduced which uses the same quadratic cost function (obtained by applying a
minimum-action argument within the framework of the Zel'dovich
approximation). In PIZA, a randomly chosen tentative correspondence between
initial and final points is successively improved by swapping {\it pairs\/} of
initial particles whenever this decreases the cost function.  Eventually, a
monotonic map is obtained which usually does not minimise the cost. This
explains the non-uniqueness of PIZA reconstruction (also noticed in
ref.~\citen{VST00}).

\begin{figure}
\iffigs 
\centerline{\psfig{file=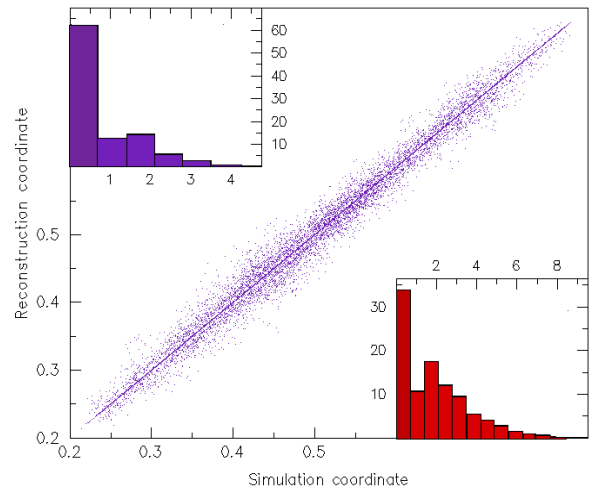,width=8cm}}
\else\drawing 65 10 {scatter}
\fi
\caption{Tests of MAK reconstructions of the Lagrangian positions, using the
data shown in Fig.~\protect\ref{f:structure}. The dots near the diagonal
are a scatter plot of reconstructed initial points vs.\ simulation initial
points for the coarsest $6.25\,h^{-1}$\,Mpc grid with 17,178 points.  The
scatter diagram uses a `quasi-periodic projection' coordinate $\tilde q \equiv
(q_1+ q_2\sqrt2 + q_3\sqrt3)/(1+\sqrt2+\sqrt3)$, which guarantees a one-to-one
correspondence between $\tilde q$-values and points on the regular Lagrangian
grid. The upper left inset is a histogram (by percentage) of distances in
reconstruction mesh units between such points; the first slightly darker bin
(whose width was taken to be slightly less than one mesh) corresponds to
perfect reconstruction (thereby allowing a good determination of the peculiar
velocities of galaxies); the lower right inset is a similar histogram for
reconstruction on a finer $3.12\,h^{-1}$\,Mpc grid using 19,187 points.  With
the $6.25h^{-1}$\,Mpc grid, 62\% of the 17,178 points are assigned perfectly
and about 75\% are within not more than one mesh.  With the $3.12h^{-1}$\,Mpc
grid, we have 34\% of exact reconstruction out of 19,187 points. On further
refinement of the mesh by a factor two, this degrades to 14\%.}
\label{f:scatter}
\end{figure}

There are, however, known deterministic strategies for the assignment problem
which give the correct unique solution; their complexity (dependence on $N$ of
the number of operations needed) is close to $N^3$ for arbitrary cost
functions, but can be sharply reduced when the cost function is
quadratic. Combining the organisation of data taken from H\'enon's mechanical
analogue machine\cite{henon95} for solving the assignment problem with the
dual simplex method of Balinski\cite{balinski86}, we have designed an
algorithm which gives the optimal assignement for about 20,000 particles in a
few hours of CPU on a fast Alpha machine. For historical reasons we call our
approach Monge--Amp\`ere--Kantorovich or MAK (see Methods).  Details of the
algorithms will be given elsewhere; we merely note that, when working with
catalogues of several hundred thousands of galaxies expected within a few
years, a direct application of the assignment algorithm in its present state
could require unreasonable computational resources.  A mixed strategy can
however be used, in which the assignment problem is solved on a coarse grid
while, on smaller scales, the Monge--Amp\`ere equation (\ref{MA}) is solved by
a relaxation technique (adapted from ref.~\citen{BB00}).

We tested the MAK reconstruction on data obtained by a cosmological $N$-body
simulation with $128^3$ particles, using the HYDRA code\cite{hydra}
(Fig.~\ref{f:structure}).  Reconstruction was performed on three $32^3$ grids
with (comoving) meshes given by $\Delta x =6.25\,h^{-1}$\,Mpc, $\Delta x/2$
and $\Delta x/4$, where $h$ is the Hubble constant in units of
100\,km\,\,s$^{-1}$\,Mpc$^{-1}$. In comoving coordinates, the typical
displacement of our mass elements over one Hubble time is about ten
$h^{-1}$\,Mpc.  We discarded those points that, at the end of the simulation
(present epoch), were not within a sphere containing about 20,000 points, a
number comparable to that of currently available all-sky galaxy redshift
catalogues. As the simulation assumes periodic boundary conditions, we also
took into account periodicity when calculating the distance between pairs of
points.  The MAK reconstructions were used to generate a scatter diagram and
various histograms allowing comparisons of simulation and reconstructed
Lagrangian points (Fig.~\ref{f:scatter}).  The results demonstrate the
essentially potential character of the Lagrangian map above $\sim
6\,h^{-1}$\,Mpc (within the given cosmological model).

We also performed PIZA reconstructions on the coarsest grid and 
obtained typically 30--40\% exactly reconstructed  points, but severe
non-uniqueness: for two different seeds of the random generator only 
about half of the exactly reconstructed points were the same.

\begin{figure}
\iffigs
\centerline{\psfig{file=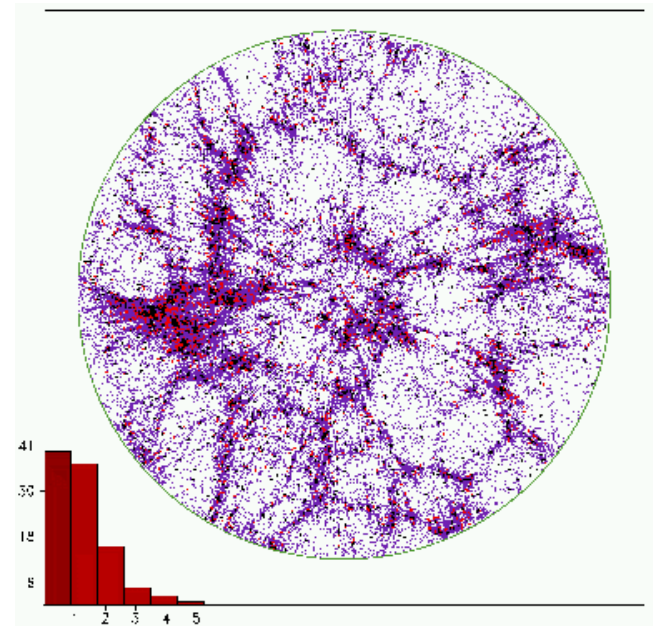,width=8.5cm}}
\else\drawing 65 10 {redshift figure here}
\fi
\caption{Reconstruction test in redshift space with the same data as for the
real-space reconstruction tested in the upper left histogram of
Fig.~\protect\ref{f:scatter}. The circular redshift map (violet points)
corresponds to the same real-space slice as displayed in
Fig.~\protect\ref{f:structure}. The observer is taken to be at the center of
the simulation box. Points used for reconstruction within the displayed slice
are highlighted in red. Reconstruction is performed by the MAK algorithm with
a different cost function, obtained (as in ref.~\citen{VST00}) by assuming
that the peculiar velocities ${\bf v}$ can be estimated by the Zel'dovich
approximation: ${\bm v}= f ({\bm \x -\q})$, where $f\approx \Omega_m^{0.6}
\approx 0.49$.  Note that we now have 43\% of exactly reconstructed points,
included among the 60\% which are within not more than $6.25\,h^{-1}$\,Mpc
from their correct positions.}
\label{f:redshift}
\end{figure}

When reconstructing from observational data, in redshift space
(Fig.~\ref{f:redshift}), the galaxies appear displaced radially (as seen by
the observer) by an amount proportional to the radial component of the
peculiar velocity. We thus performed another reconstruction, with an
accordingly modified cost function, that led to somewhat degraded
results (Fig.~\ref{f:redshift}) but at the
same time provided an approximate determination of peculiar velocities.
More accurate determination of peculiar velocities
can be done using second-order Lagrangian perturbation theory. The effect of
the catalogue selection function can be handled by standard techniques; for
instance one can assign each galaxy a `mass' inversely proportional to the
catalog selection function \cite{VST00,BEN01}.

What is the smallest length scale at which an optimisation algorithm such as
MAK can be expected to give a unique and reliable reconstruction? The key
ingredient here is the {\it simultaneous\/} knowledge of the initial and
present mass density fields. MAK-type reconstruction (with a suitable cost
based on the equation of a self-gravitating fluid) should therefore be 
possible down to
scales comparable to the thickness of collapsed structures, below which the
hydrodynamical description ceases to be meaningful.

The fact that MAK guarantees a unique solution and that our present
reconstruction hypothesis proved to be very faithful down to
$6.5\,h^{-1}\,$Mpc makes our method very promising for the analysis of
galaxy redshift surveys\cite{survey}.  Reconstruction of the primordial positions
and velocities of matter will allow us to test the Gaussian nature of
the primordial perturbations and the self-consistency of cosmological
hypotheses, such as the choice of the global cosmological parameters
and the assumed biasing scheme.  By obtaining a point-by-point
reconstruction of the specific realisation that describes the observed
patch of our Universe, we can distinguish between
universal properties and the influence of the large-scale environment
on the galaxy formation process.  Moreover, reconstruction will open a
new window not only onto the past but also into the present Universe:
it would enable us to make a first-time determination of the peculiar
velocities of a very large number of galaxies, using their positions in 
redshift catalogues.

\vspace*{0.3cm}
{\footnotesize
\par\noindent {\bf Methods}
\vspace*{0.2cm}
\par\noindent {\bf Monge--Amp\`ere equation}
\vspace*{0.1cm}

\noindent  The Lagrangian map $\q\mapsto \x$ is taken to be the gradient of a
convex potential $\Phi(\q)$; therefore its inverse $\x\mapsto \q$ also has a
potential representation $\q =\nabla_{\x}\Theta(\x)$, where $\Theta(\x)$ is
again a convex function; the two potentials are Legendre--Fenchel transforms of
each other (see ref.~\citen{arnold}, pp.~61--65):
\begin{equation}
\!\!\Theta(\x) = \max_\q\,\left[ \q\cdot\x -\Phi(\q)\right] ;
\!\!\!\! \quad \Phi(\q) =
\max_\x\, \left[\x\cdot\q -\Theta(\x)\right].
\label{legendre}
\end{equation}
The potential $\Theta$ satisfies the Monge--Amp\`ere equation (\ref{MA}),
written for the first time by Amp\`ere\cite{ampere20} by exploiting the
property of the Legendre transformation. Note that within the more restricted
framework of the Zel'dovich approximation, $\Theta$ differs just by a
quadratic additive term from the Eulerian velocity potential\cite{VDFN94}.

\vspace*{0.3cm}
\par\noindent {\bf Quadratic cost function}
\vspace*{0.1cm}

\noindent To show that the quadratic cost minimisation leads to the
Monge--Amp\`ere equation, we define the displacement field ${\bm \xi}(\x)
\equiv \x-\q(\x)$ and perform a variation $\delta {\bm \xi}(\x)$ to obtain,
to lowest order, the variation of the cost function $ \,\delta I \,=
\,\int_\x 2 \,{\bm
\xi}(\x)\cdot(\rho(\x)\,\delta {\bm \xi}) \, d^3x$.  
\vbox{\noindent The condition that the
Eulerian density remains unchanged, which constrains the variation, is
expressed as $\nabla_\x \cdot (\rho(\x) \delta {\bm \xi}(\x)) = 0$. By a
simple Lagrange multiplier argument, this implies that ${\bm \xi}$ must be a
gradient of some function of $\x$; thus, $\q = \x - {\bm \xi}=\nabla_\x
\Theta(\x)$.  Furthermore, should $\Theta$ be
non-convex and thus lead to multi-streaming, this would prevent the
Lagrangian map from being optimal.
\vspace*{0.3cm}
\par\noindent {\bf History of mass transportation}
\vspace*{0.1cm}

\noindent Monge\cite{monge81} posed the following problem: how to optimally
move material from one place to another, knowing only its initial and final
spatial distributions, the cost being a prescribed function of the distance
travelled by `molecules' of material (a linear function in Monge's original
work).  Kantorovich\cite{kantorovich42} showed that Monge's query was an
instance of the linear programming problem and developed for it a theory which
found numerous applications in economics and applied mathematics. 
}
}

\vspace*{-2mm}
\par\noindent {\bf Acknowledgements}
\vspace{1mm}

\noindent Special thanks are due to E.~Branchini (observational and conceptual
aspects), to Y.~Brenier (mathematical aspects) and to M.~H\'enon (algorithmic
aspects and the handling of spatial periodicity and of scatter plots).  We
also thank J.~Bec, H.~Frisch, B.~Gladman, L.~Moscardini, A.~Noullez,
C.~Porciani, M.~Rees, E.~Spiegel, A.~Starobinsky and P.~Thomas for comments.
This work was supported by the BQR program of Observatoire de la C\^ote
d'Azur, by the TMR program of the European Union (UF, RM), by MIUR (SM), by
the French Ministry of Education, the McDonnel Foundation, the Russian RFBR
and INTAS (AS).

\end{document}
\vspace*{2mm}
\noindent
Correspondence and requests for materials should be addressed to U.~Frisch
(e-mail: uriel@obs-nice.fr).
\end{document}